%% file: main.tex
\documentclass[sigconf]{acmart}

\usepackage{hyperref}
\usepackage{hyperxmp}
\usepackage{xcolor}
\usepackage{tikz}
\usepackage{courier}
\usepackage{tabularx}
\usepackage{makecell}
\usepackage{url}
\usepackage[most]{tcolorbox}
\tcbuselibrary{listingsutf8}

\newtcolorbox[auto counter, number within=section]{mybox}[2][]{%
  float,
  colback=white,
  colframe=black!60,
  fonttitle=\bfseries,
  enhanced,
  breakable,
  sharp corners=south,
  boxrule=0.5pt,
  left=1mm, right=1mm, top=1mm, bottom=1mm,
  title=Box~\thetcbcounter: ~#2,
  #1
}

\definecolor{greenreference}{HTML}{45b445}
\input{meta}  

\copyrightyear{2025}
\acmYear{2025}
\setcopyright{acmlicensed}\acmConference[ICAIF '25]{6th ACM International Conference on AI in Finance}{November 15--18, 2025}{Singapore, Singapore}
\acmBooktitle{6th ACM International Conference on AI in Finance (ICAIF '25), November 15--18, 2025, Singapore, Singapore}
\acmDOI{10.1145/3768292.3770363}
\acmISBN{979-8-4007-2220-2/2025/11}

\input{sections/abstract}
\begin{document}

    \maketitle

\input{sections/introduction}

\input{sections/related_work}

\input{sections/problem_statement}
\input{sections/methodology}

\input{sections/evaluation_metrics}

\input{sections/discussion}
\input{sections/conclusion}

\bibliographystyle{ACM-Reference-Format}

\input{references.tex}

\end{document}

%% file: meta.tex
\title{FinReflectKG: Agentic Construction and Evaluation of Financial Knowledge Graphs}

\author{Abhinav Arun}
\affiliation{%
  \institution{Domyn}
  \city{New York}
  \country{US}
}
\email{abhinav.arun@domyn.com}

\author{Fabrizio Dimino}
\affiliation{%
  \institution{Domyn}
  \city{New York}
  \country{US}
}
\email{fabrizio.dimino@domyn.com}

\author{Tejas Prakash Agarwal}
\affiliation{%
  \institution{Domyn}
  \city{New York}
  \country{US}
}
\email{tejas.p.agrawal@domyn.com}

\author{Bhaskarjit Sarmah}
\affiliation{%
  \institution{Domyn}
  \city{Gurgaon}
  \country{India}
}
\email{bhaskarjit.sarmah@domyn.com}

\author{Stefano Pasquali}
\affiliation{%
  \institution{Domyn}
  \city{New York}
  \country{US}
}
\email{stefano.pasquali@domyn.com}

%% file: sections/abstract.tex
\begin{abstract}
The financial domain poses unique challenges for knowledge graph (KG) construction at scale due to the complexity and regulatory nature of financial documents. Despite the critical importance of structured financial knowledge, the field lacks large-scale, open-source datasets capturing rich semantic relationships from corporate disclosures. We introduce an \textbf{open-source, large-scale financial knowledge graph dataset built from the latest annual SEC 10-K filings of all S\&P 100 companies} - a comprehensive resource designed to catalyze research in financial AI.

We propose a robust and generalizable knowledge graph (KG) construction framework that integrates intelligent document parsing, table-aware chunking, and schema-guided iterative extraction with a reflection-driven feedback loop. Our system incorporates a comprehensive evaluation pipeline, combining rule-based checks, statistical validation, and LLM-as-a-Judge assessments to holistically measure extraction quality. We support three extraction modes-single-pass, multi-pass, and reflection-agent-based allowing flexible trade-offs between efficiency, accuracy, and reliability based on user requirements. Empirical evaluations demonstrate that the reflection-agent-based mode consistently achieves the best balance, attaining a 64.8\% compliance score against all rule-based policies (CheckRules) and outperforming baseline methods (single-pass \& multi-pass) across key metrics such as precision, comprehensiveness, and relevance in LLM-guided evaluations.

The utility of our KG pipeline is demonstrated through its flexible extraction modes, coupled with a multi-faceted evaluation methodology. By releasing a high-quality, thoroughly evaluated dataset along with a comprehensive KG construction \& evaluation framework, we aim to advance transparency, reproducibility, and innovation in financial KG research. \textbf{The dataset is publicly available at:} \url{https://anonymous.4open.science/r/KG-Financial-Datasets-SP-100-529B/README.md}

\end{abstract}

%% file: sections/introduction.tex
\section{Introduction}

Knowledge graphs (KGs) have become foundational for organizing and reasoning over complex, interconnected information in domains ranging from biomedicine to finance~\cite{benetka2017financial, elhammadi2020pipeline, li2024findkg}. In the financial sector, the construction of high-quality KGs is particularly challenging due to the heterogeneous and highly interconnected nature of documents such as SEC 10-K filings. Although recent advances in large language models (LLMs) have enabled significant progress in information extraction~\cite{li2023evaluating, zhang2024extract}, there is a general lack of reliable KG benchmark datasets for the financial domain \cite{koncel2023bizbench, reddy2024docfinqa}, which hinders the widespread adoption of KGs for downstream financial applications. Additionally, most existing financial KGs are either limited in scope, relying primarily on news feeds ~\cite{li2024findkg, elhammadi2020pipeline}, or lack the rigorous evaluation required for building trust for wide scale deployment \& adoption within the financial industry \cite{kertkeidkachorn2023finkg}.

In this work, we address the gaps by introducing a large-scale financial KG dataset built exclusively using annual SEC 10-K filings of all the S\&P 100 companies for the year 2024. Our approach is inspired by recent developments in prompt-driven and iterative extraction~\cite{carta2023iterative, jiang2025ras, wei2023chatie}, schema canonicalization~\cite{zhang2024extract}, and self-reflective LLM agents~\cite{li2024findkg}. We have built a robust pipeline that combines intelligent document parsing, table-aware chunking, schema-guided iterative extraction, and reflection-driven feedback, culminating in a thoroughly evaluated resource for the financial AI community.

Our framework supports three extraction modes—single-pass, multi-pass, and reflection-agent-based and we further introduce a dynamic schema configuration process, leveraging both LLMs and domain experts to ensure business relevance and adaptability. The resulting dataset and methodology empower a range of downstream applications, including KG-powered entity search, multi-hop question answering, signal generation via news integration, and advanced graph powered predictive models \& network analytics.

The main contributions of our paper are as follows:
\begin{itemize}
\item \textbf{Large-Scale Open Source Financial KG Dataset:} We release an open-source, comprehensive financial knowledge graph constructed from SEC 10-K filings of all S\&P 100 companies, providing a massive, high-quality resource for financial AI research and applications.

\item \textbf{Novel Reflection Driven Extraction Framework:} We introduce a three-mode extraction pipeline (single-pass, multi-pass, reflection-agent-based) with reflection driven feedback systematically improving extraction quality through iterative refinement, achieving 64.8\% compliance across all CheckRules and best voted mode in the LLM-as-a-Judge evaluation.

\item \textbf{Generalizable Evaluation Methodology:} We develop a holistic evaluation framework encompassing rule-based compliance checks, coverage analysis, semantic diversity metrics, and LLM-as-a-Judge comparative assessment, establishing new benchmarks for financial KG evaluation.
\end{itemize}

By releasing this high-quality, rigorously evaluated dataset and pipeline, we aim to advance transparency, reproducibility, and innovation in financial knowledge graph research and its applications.

%% file: sections/related_work.tex
\section{Related Work}
In this section, we provide details on existing research that has been happening in the realm of Knowledge Graph Construction with a focus on the financial domain. 

\textbf{Knowledge Graph Construction:} 
Knowledge graph construction (KGC) has traditionally followed a pipeline architecture composed of discrete subtasks such as entity recognition \cite{zukov2018neural, martins2019joint} and relation classification \cite{zeng2014relation, zeng2015distant}, often relying on supervised learning with domain-specific annotations. However, the recent success of LLMs in information extraction tasks \cite{li2023evaluating} highlights a paradigm shift toward prompt-based strategies for robust knowledge graph construction.

Recent work has leveraged large language models (LLMs) for zero and few-shot extraction using iterative prompting strategies. \citet{carta2023iterative} propose a zero-shot pipeline using GPT-3.5, where a sequence of prompts progressively identifies entities, types them, extracts relations, and resolves co-references without supervision or external knowledge bases. \citet{jiang2025ras} introduce the Retrieval-And-Structuring (RAS) framework, which alternates between query planning, retrieval, and triple extraction to incrementally construct task-specific mini-KGs, outperforming standard RAG methods. These approaches complement interactive designs like ChatIE \cite{wei2023chatie}, which frames triple extraction as a multi-turn QA task. Collectively, they underscore the emergence of modular, prompt-driven workflows for scalable KG construction with LLMs.

\textbf{Normalization and Schema Canonicalization} : 
While open information extraction (OIE) methods enable large-scale triples extraction, they often produce unstandardized and semantically redundant outputs. Without canonicalization, multiple surface-level variations of the same relation (e.g., \textit{supplies}, \textit{is supplier of}) can coexist in the knowledge graph, introducing ambiguity and reducing its utility for downstream tasks.
Traditional approaches depend on whether a target schema is available: alignment-based methods use lexical resources like WordNet, while schema-free methods such as CESI \cite{vashishth2018cesi} cluster relations using embeddings and external signals. However, clustering often over-generalizes, merging distinct semantics. The recent Extract-Define-Canonicalize (EDC) framework \cite{zhang2024extract} offers a more robust, LLM-native alternative. It extracts triples, defines schema candidates in context, and canonicalizes them using LLM-generated definitions and reasoning-avoiding brittle heuristics or external ontologies.

\textbf{Financial Knowledge Graphs:}
Early work on financial knowledge graphs (KGs) focused on extracting structured event representations from unstructured news. \citet{benetka2017financial} jointly extract all attributes of economic transactions as quintuples, aggregating information across multiple mentions to build unified event-centric graphs. Subsequent efforts, such as \citet{elhammadi2020pipeline}, combined semantic role labeling, dependency parsing, and domain-specific dictionaries to construct high-precision financial KGs from news. More recently, \citet{li2024findkg} introduced FinDKG, a dynamic, time-varying financial KG generated from news using a fine-tuned LLM pipeline, systematically extracting entities and relations as event quadruples. Collectively, these works highlight the evolution from rule-based and supervised pipelines to LLM-driven, schema-flexible approaches for building robust financial knowledge graphs.

Building upon these foundations, we present an open-source, large-scale financial knowledge graph constructed exclusively from SEC 10-K filings of all S\&P 100 companies. Our work addresses key limitations in existing approaches: while previous financial KGs focused on news-based event extraction, we target the most authoritative financial documents (SEC filings) with comprehensive schema-guided extraction, establishing new benchmarks for financial KG construction and evaluation.

%% file: sections/problem_statement.tex
\section{Problem Formulation and Schema Design}
Building on recent advances in large language models and prompt engineering, we design a pipeline to extract high-quality financial knowledge graph triples of the form (Head Entity, Head Type, Relationship, Tail Entity, Tail Type) from the SEC 10-K filings. 


\begin{definition}
A \emph{triple} in our context is a 5-tuple of the form\\
\noindent\hspace*{1em}%
\texttt{(Head Entity,\ Head Type,\ Relationship,\ Tail Entity,\ Tail Type)},\\
where the Head and Tail Entities are linked by a semantic Relationship, and each entity is annotated with its type. Throughout this paper, we refer to these 5-tuples as triples.
\end{definition}

Our pipeline begins with a business-driven schema configuration approach (closed information extraction), where entity types and relationships are primarily defined by business subject matter experts (SMEs), specific to input financial data feeds and downstream applications \cite{kertkeidkachorn2023finkg, li2024findkg}. For instance, the same SEC 10-K filing can be utilized differently by portfolio managers and risk managers. We employ an AI-assisted reconciliation approach where we prompt the underlying LLM to propose a schema given sample documents, but the schemas are ultimately approved and defined by SMEs. This approach ensures domain relevance and business alignment.

We focus on extracting KG triples in a closed information extraction setting, which leads to less noisy knowledge graphs and enables the construction of reliable KGs that can be directly integrated into downstream applications. For SEC filings, we have created a comprehensive schema through collaboration between LLMs and financial SMEs, a subset of which is presented in Tables \ref{tab:entity_types} and \ref{tab:relationships}. This schema captures the complex relationships and entities specific to financial reporting and regulatory compliance.

\begin{table}[htbp]
\small
\renewcommand{\arraystretch}{1.2}
\centering
\caption{Subset of Pre-Configured Entity Types and their Definitions for Financial KG Construction}
\label{tab:entity_types}
\begin{tabularx}{\columnwidth}{|>{\hsize=0.3\hsize\raggedright\arraybackslash}X|>{\hsize=0.7\hsize\raggedright\arraybackslash}X|}
\hline
\textbf{Entity Type} & \textbf{Definition} \\
\hline
ORG & Filing Company (Issuer: The public company that is the subject of the 10-K filing) \\
\hline
PERSON & Key individuals (e.g., CEO, CFO, Board members) \\
\hline
COMP & External companies referenced in the filing, including competitors, suppliers, customers, or partners \\
\hline
PRODUCT & Products or services offered by the company or competitors (e.g., iPhone, AWS) \\
\hline
SEGMENT & Internal divisions or business segments of the filer ORG (e.g., Cloud segment, North America retail) \\
\hline
FIN\_METRIC & Financial metrics or values (e.g., Net Income, EBITDA, CapEx, Revenue) \\
\hline
RISK\_FACTOR & Documented risks (e.g., market risk, supply chain risk, regulatory risk) \\
\hline
EVENT & Material events such as pandemics, natural disasters, M\&A events, regulatory changes \\
\hline
REGULATORY\_ REQUIREMENT & Specific regulations or legal frameworks (e.g., Basel III, GDPR, SEC requirements) \\
\hline
ESG\_TOPIC & Environmental, Social, and Governance themes (e.g., Carbon Emissions, DEI, Climate Risk) \\
\hline
\end{tabularx}
\end{table}

\begin{table}[htbp]
\small
\centering
\renewcommand{\arraystretch}{1.2}
\caption{Subset of Pre-Configured Relationship Types and their Definitions for Financial KG Construction}
\label{tab:relationships}
\begin{tabularx}{\columnwidth}{|l|X|}
\hline
\textbf{Rel. Type} & \textbf{Definition} \\
\hline
Has\_Stake\_In & Indicates full or partial ownership or equity interest \\
\hline
Operates\_In & Indicates operational geography or market presence \\
\hline
Produces & Manufactures or develops a product or service \\
\hline
Impacts & Specifies the broad influence or effect an entity or event has on financial performance, market trends, or other key outcomes \\
\hline
Involved\_In & Specifies direct involvement in an event such as a merger, acquisition, or litigation \\
\hline
Impacted\_By & Indicates that the entity was materially affected by a major event \\
\hline
Discloses & Reveals or reports information, metrics, or developments \\
\hline
Complies\_With & Meets regulatory or policy requirements \\
\hline
Supplies & Indicates vendor or supplier relationship \\
\hline
Partners\_With & Indicates formal or strategic collaboration \\
\hline
\end{tabularx}
\end{table}

%% file: sections/methodology.tex
\section{Methodology}
\begin{figure}[htbp]
    \centering
    \includegraphics[width=\columnwidth]{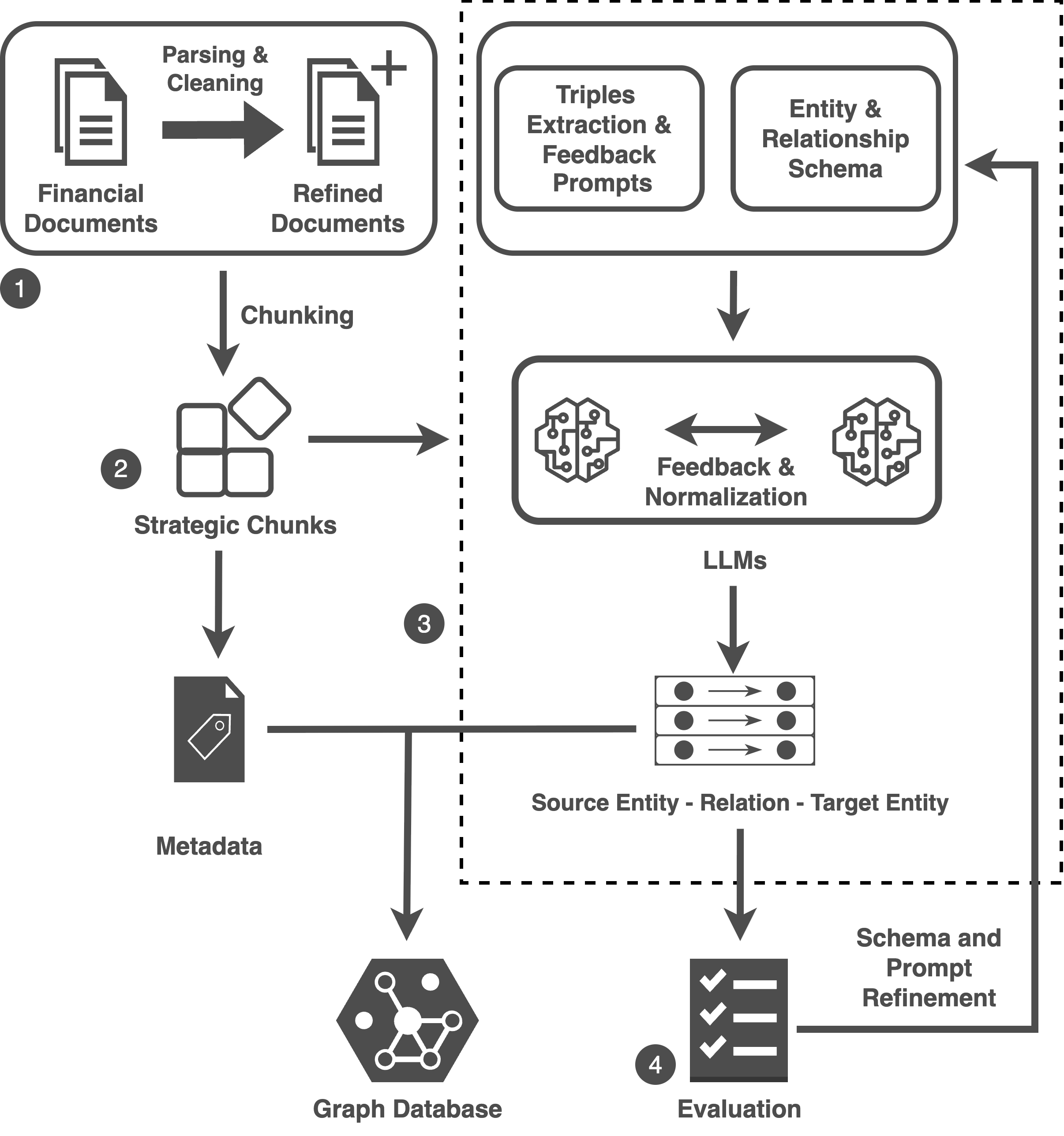}
    \caption{Overview of the Graph RAG microservice and its integration with the KG construction pipeline.}
    \label{fig:graph_rag_microservice}
\end{figure}

Once we have configured the schema \& defined the ontology, our KG construction pipeline comprises of four interlocking components: (1) Intelligent Document Parsing Layer, (2) Table-Aware Semantic Chunking Layer, (3) Iterative Prompt \& Agent Driven Triples Extraction Layer, and (4) Robust Evaluation Layer. By integrating best practices from AI-driven information extraction and financial KG construction, we achieve both high precision and domain relevance. The overall pipeline is showcased in the Figure~\ref{fig:graph_rag_microservice}

\subsection{Intelligent Document Parsing Layer}

Our pipeline leverages an advanced document parsing layer built on \texttt{docling} \cite{auer2024docling}, enabling robust extraction and retention of the diverse formats present in SEC 10-K filings. This layer preserves the multiformat architecture of the source documents, including narrative text, tables, and images. Tables are especially critical in financial filings, as they often encapsulate key quantitative disclosures (e.g., revenue by segment, risk breakdowns, or financial metrics) essential for downstream knowledge graph construction.

The parser operates in two modes: a \textbf{multimodal mode}, which combines OCR and image annotation to capture both textual and visual information, and a \textbf{text-only mode} for efficient processing when only narrative content is required. For the scope of this paper, we used only the text mode. In both modes, tables structures are retained as markdown context, maintaining row–column associations and semantic context. Narrative sections are tagged with section headers (e.g., ``Risk Factors,'' ``Management's Discussion'') to facilitate schema alignment in subsequent pipeline stages.

This intelligent parsing approach ensures that no critical information is lost during preprocessing, and provides a high-fidelity, semantically annotated representation of the original document for downstream chunking and extraction.

\subsection{Table-Aware Semantic Chunking Layer}

To accommodate the token limits of modern large language models, our pipeline incorporates a custom chunking algorithm that strategically segments documents while preserving semantic and structural integrity. Unlike naive sliding window or fixed-length approaches, our chunker is \textbf{table-aware}: any detected table in the markdown is retained as a single atomic chunk, ensuring that row and column relationships—and thus the full context of financial data—are never fragmented. This is particularly important for SEC 10-K filings, where tables often encode critical quantitative disclosures.

We employ a \textbf{section-aware text segmentation}, splitting text at logical boundaries such as paragraphs or subsection headings to maintain topical coherence. Each chunk is constrained to a maximum size of 2048 tokens (\texttt{CHUNK\_SIZE = 2048}), balancing context preservation with LLM input requirements.

This approach ensures that both tabular and textual information are optimally prepared for downstream extraction, maximizing the fidelity and relevance of the knowledge graph construction process.

\subsection{Iterative Prompt \& Agent Driven Triples Extraction Layer}
Leveraging the predefined schema as highlighted in Tables \ref{tab:entity_types} and \ref{tab:relationships}, we adopt an iterative empirical approach to identify the paradigm that gives us the most reliable and grounded KG triples. We have used Qwen2.5-72B-Instruct as the LLM for KG construction.
The prompts explicitly enumerate these categories to constrain LLM outputs.

\subsubsection{Single–Pass Workflow}
In the single-pass mode, we employ a single, comprehensive prompt that instructs the language model to extract all valid knowledge graph triples from each document chunk in one step. The prompt enforces the use of only pre-defined entity and relation types, requires normalization of entity names (e.g., mapping all company references to the ticker), and outputs the results in a strict JSON format for downstream processing. While efficient, this approach may still yield occasional inconsistencies in entity normalization or relation assignment due to the inherent limitations of single-turn LLM prompting.
Mathematically, for each chunk $c$ and schema $S$:


\begin{equation}
T_c^{(1)} = \mathrm{Extract}(c, S, \phi_{\mathrm{sp}})
\label{eq:single_pass_prompting}
\end{equation}

where $T_c^{(1)}$ is the set of extracted and normalized triples for chunk $c$, and $\phi_{\mathrm{sp}}$ denotes the single pass (sp) mode (prompts).
\subsubsection{Multi–Pass Workflow}

To improve extraction quality and consistency, we adopt a multi-pass prompting strategy. In this approach, the language model first extracts candidate triples from each chunk using the pre-defined schema. In a second pass, the model re-ingests its own output alongside the original chunk and applies a dedicated normalization prompt to:
\begin{itemize}
  \item Enforce canonical naming (e.g., ticker substitution for company references).
  \item Filter to schema-compliant entity and relation types.
  \item Merge duplicate or redundant entities and relationships.
  \item Validate directionality and ordering for all relations.
  \item Remove or correct invalid or ambiguous triples.
\end{itemize}
This two-step process leverages the LLM's reasoning capabilities for both extraction and refinement, resulting in higher precision and more consistent knowledge graph triples.
Mathematically, for each chunk $c$ and schema $S$:

\begin{equation}
T_c^{(1)} = \mathrm{Extract}(c, S, \phi_{\mathrm{mp}})
\label{eq:multi_pass_prompting_1}
\end{equation}

\begin{equation}
T_c^{(2)} = \mathrm{Normalize}(c, T_c^{(1)}, S, \phi_{\mathrm{mp}})
\label{eq:multi_pass_prompting_2}
\end{equation}

where $T_c^{(1)}$ is the initial set of extracted triples, $T_c^{(2)}$ is the refined, schema-compliant set after normalization, and $\phi_{\mathrm{mp}}$ denotes the multi pass (mp) mode (prompts).

\subsubsection{Reflection‐Driven Agentic Workflow and Meta‐Analysis}
We deploy a dedicated reflection agent that iteratively refines the initial set of triples by simulating a multi turn interaction between the feedback (critic) \& correction LLM.
The critic LLM specifically: 
\begin{itemize}
  \item Verifies the entity labels and relation assignments against the domain schema.
  \item Assesses business relevance and flags low‐value or contradictory triples.
\end{itemize}
Feedback is returned in a structured JSON schema enabling automated ingestion as shown in box ~\ref{box:feedback-llm-sample}. All critique instances are logged for meta‐analysis, revealing recurrent error patterns and informing prompt redesign. Our reflection approach is inspired by recent advances in self-reflective and memory-augmented language agents~\cite{li2024reflexion,shinn2023reflexion}.\\

\begin{mybox}[label={box:feedback-llm-sample}]{Sample Response from the Feedback LLM}
\footnotesize
\textbf{\texttt{\{}
\\ \hspace{2mm} "triple\_number": "Triple N",
\\ \hspace{2mm} "triple": [\textcolor{red}{"We"}, "ORG", "Impacted\_By", "supply chain disruptions", \textcolor{red}{"RISK\_TYPE"}],
\\ \hspace{2mm} "issue": \textcolor{red}{"Indirect reference to an entity in the triple. RISK\_TYPE is not a valid preconfigured category"},
\\ \hspace{2mm} "suggestion": \textcolor{greenreference}{"replace We with NVDA as this information comes from Nvidia's 10-K file; substitute RISK\_TYPE with RISK\_FACTOR from the configured entity types"}
\\ \texttt{\}}}
\end{mybox}


Let $T_c^{(1)}$ be the initial set of triples for chunk $c$ (from the extraction LLM), $S$ is the preconfigured schema and $\phi_{\mathrm{re}}$ denotes the reflection (re) extraction mode. For each reflection step $t = 1, 2, \ldots, n$:
\begin{itemize}
\item \textbf{Extraction LLM} generates an initial set of triples $T_c^{(1)}$ which are further validated and refined in a cyclic loop of feedback \& correction LLMs.
\item \textbf{Feedback LLM} analyzes $T_c^{(t-1)}$ and produces a set of issues $F_c^{(t)}$ and suggestions.
\item \textbf{Correction LLM} updates problematic triples (or drops them) to produce $T_c^{(t)}$.
\end{itemize}
\begin{equation}
T_c^{(1)} = \mathrm{Extract}(c, S, \phi_{\mathrm{re}})
\label{eq:reflection_prompting_1}
\end{equation}

\begin{equation}
F_c^{(t)} = \mathrm{Feedback}(c, T_c^{(t-1)}, S, \phi_{\mathrm{re}}) \\
\label{eq:reflection_prompting_2}
\end{equation}

\begin{equation}
T_c^{(t)} = \mathrm{Correct}(c, T_c^{(t-1)}, F_c^{(t)}, S, \phi_{\mathrm{re}})
\label{eq:reflection_prompting_3}
\end{equation}

where $\phi_{\mathrm{re}}$ denotes the reflection (re) mode.\\
\textbf{Stopping criteria:} \\
Stop at step $t^*$ if $F_c^{(t^*)} = \emptyset$ (no issues found) or $t^* = n_{\text{max}}$ (max steps reached which can be empirically determined for different LLMs). This is intelligently determined using the agentic prowess of the underlying LLMs.\\
\textbf{Final output:}
\begin{equation}
T_c^{(*)} = T_c^{(t^*)}
\label{eq:final_output_1}
\end{equation}

For all chunks in document $D$:
\begin{equation}
T_D^{(*)} = \bigcup_{i=1}^N T_{c_i}^{(*)}
\label{eq:final_output_2}
\end{equation}

\begin{figure}[htbp]
    \centering
    \includegraphics[width=\columnwidth]{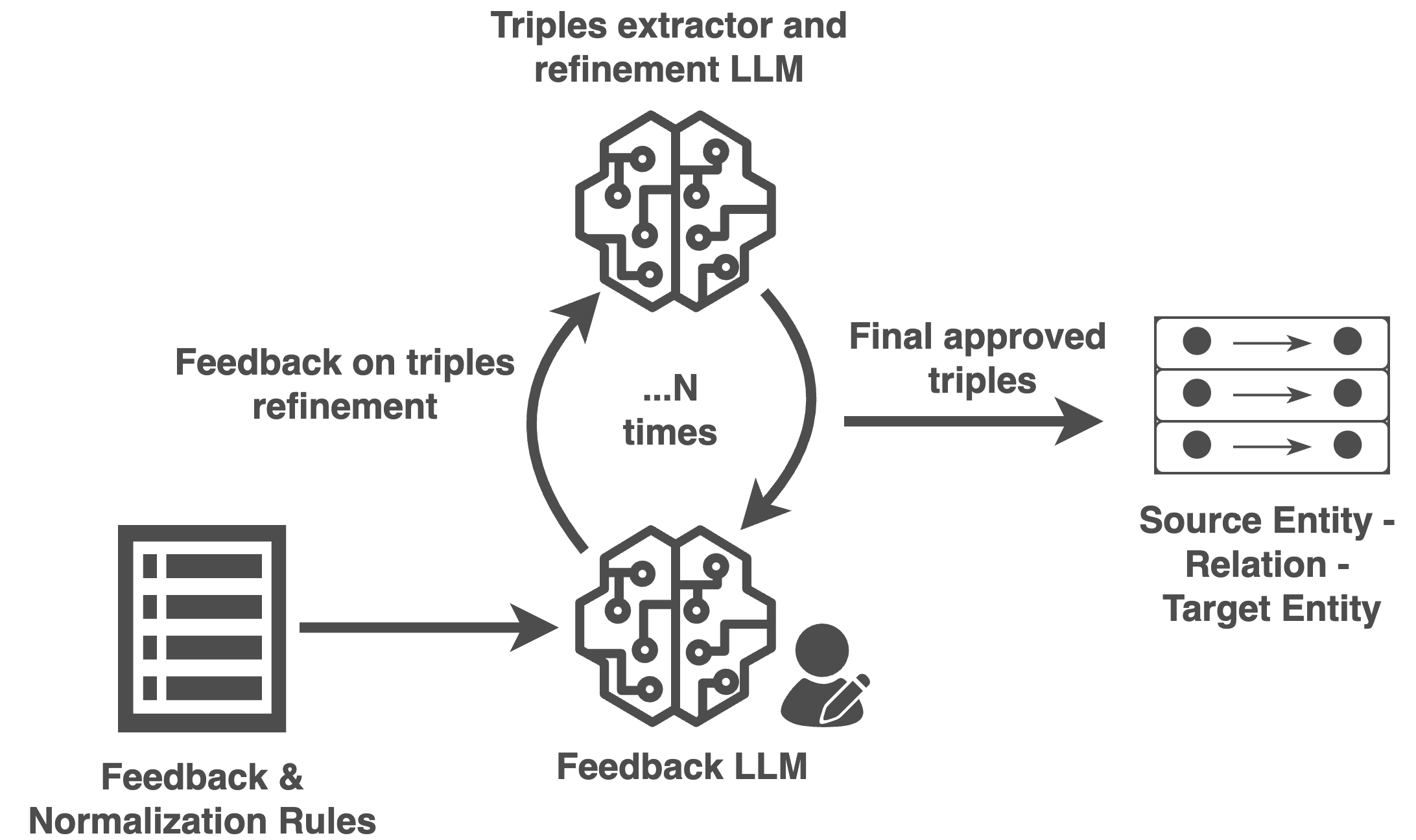}
    \caption{Overview of Reflection Agent where the triples are refined using an iterative feedback loop}
    \label{fig:reflection_paradigm}
\end{figure}

This iterative prompt engineering \& agentic approach advances the current state of domain‐specific KG extraction and offers a reusable framework for other regulated domains.

%% file: sections/evaluation_metrics.tex
\section{Evaluation} 
Evaluating knowledge graph construction requires overcoming limitations of traditional metrics that rely on simplistic or single-dimensional measures, especially when comprehensive ground-truth annotations are unavailable \citep{choi2025kgc}. To address this, we propose a holistic evaluation framework integrating complementary methodologies:

\subsection{CheckRules}
Knowledge graph construction systems often suffer from inconsistent entity normalization and schema violations, leading to redundant representations and semantic ambiguity. A critical issue arises from abstract entity references that lack semantic clarity: pronouns like "the company", "we", "our", or "it" create ambiguous nodes that cannot be properly linked or reasoned upon. For instance, in a financial document about Apple Inc., the same corporate entity may be represented as "Apple Inc.", "the company", "we", or the standardized ticker "AAPL", creating artificial multiplicity and semantic confusion that degrades graph quality and downstream reasoning capabilities.

To address these systematic issues, \textbf{CheckRules} evaluates each extracted triple against a set of rules:
\begin{itemize}
    \item \textbf{Subject Reference}: Identifies and flags abstract entity references (e.g., "the company", "we", "our", "it") that lack semantic specificity (e.g., "AAPL" for Apple Inc.).
    \item \textbf{Entity Length Constraint}: Limits entity names to maximum 5 words to prevent verbose representations.
    \item \textbf{Entity Schema Compliance}: Ensures extracted entities types adhere to a predefined schema.
    \item \textbf{Relationship Schema Compliance}: Ensures extracted relationship adhere to a predefined schema.
\end{itemize}

Each extracted triple is individually evaluated against these rules. For a triple \(t\) with \(R\) rules, the CheckRules score is:
\begin{equation}
    CR(t)=\frac{1}{R}\sum_{i=1}^{R}\phi_i(t)
\label{eq:check_rules}
\end{equation}
where \(\phi_i(t)\in\{0,1\}\) indicates compliance. 

\begin{table}[h]
\centering
\caption{CheckRules scores for different extraction modes }
\label{tab:checkrules_individual}
\begin{tabularx}{\columnwidth}{l|XXX}
\hline
\textbf{Rule(\textuparrow)} & \small \textbf{Single Pass (\%)} & \small \textbf{Multi Pass (\%)} & \small \textbf{Reflection (\%)} \\
\hline
Subject Reference & 99.9 & \textbf{100.0} & \textbf{100.0} \\
Entity Length Constraint & 68.2 & \textbf{79.5} & 78.0 \\
Entity Schema Compliance & 95.9 & 96.6 & \textbf{98.1} \\
Rel. Schema Compliance & 64.6 & 62.0 & \textbf{84.2} \\
\hline
\end{tabularx}
\end{table}

\begin{table}[h]
\centering
\caption{Proportion of valid triples with increasingly strict criteria}
\label{tab:checkrules_aggregate}
\begin{tabularx}{\columnwidth}{l|XXX}
\hline
\textbf{Metric (\textuparrow)} & \textbf{Single Pass (\%)} & \textbf{Multi Pass (\%)} & \textbf{Reflection (\%)} \\
\hline
At least 1 rule & 100.0 & 100.0 & 100.0 \\
At least 2 rules & 99.1 & 99.4 & \textbf{99.8} \\
At least 3 rules & 87.3 & 90.8 & \textbf{95.6} \\
At least 4 rules & 42.3 & 47.3 & \textbf{64.8} \\
\hline
\end{tabularx}
\end{table}

While \autoref{tab:checkrules_individual} identifies specific compliance bottlenecks, \autoref{tab:checkrules_aggregate} quantifies the proportion of triples considered valid under increasingly strict criteria. The results show that company references and entity-type compliance are consistently handled across all modes. However, relationship-schema compliance lags behind, indicating the need for further refinements of the predefined schema\cite{wadhwa2023revisiting}. Notably, the Reflection paradigm substantially improves compliance, particularly concerning entity naming length and relationship schema rules. \\
The second table indeed confirms this by highlighting Reflection's notable performance advantage at higher compliance thresholds. 

\subsection{Local Extraction Efficiency}

To assess extraction effectiveness and comprehensiveness, we compute Coverage Ratios, quantifying diversity and completeness across entities, entity types, and relationships. Specifically, we measure the proportion of unique entities and entity types relative to total extracted elements (Entity Coverage Ratio, ECR; Type Coverage Ratio, TCR), and evaluate schema utilization through normalized coverage ratios (TCR-N, RCR-N). Similar computations for relationships yield Relationship Coverage Ratios (RCR, RCR-N). 

\begin{table}[htbp]
\centering
\caption{Local Extraction Efficiency scores across different extraction modes}
\label{tab:coverage_assessment}
\begin{tabularx}{\columnwidth}{l|XXX}
\hline
\textbf{Metric (\textuparrow)} & \textbf{Single Pass (\%)} & \textbf{Multi Pass (\%)} & \textbf{Reflection (\%)} \\
\hline
Triples (per chunk) & 13.3 & 12.4 & \textbf{15.8} \\
ECR & 0.30 & 0.31 & \textbf{0.53} \\
TCR & 0.09 & 0.10 & \textbf{0.27} \\
TCR-N & 0.13 & 0.13 & \textbf{0.18} \\
RCR & 0.21 & 0.22 & \textbf{0.38} \\
RCR-N & 0.13 & 0.13 & \textbf{0.14} \\
\hline
\end{tabularx}

\end{table}

\autoref{tab:coverage_assessment} highlights a clear performance hierarchy, with Reflection consistently outperforming Single Pass and Multi Pass across all coverage metrics. Reflection generates more triples per chunk and achieves substantially higher entity, type, and relationship coverage, indicating a richer and more diverse extraction of semantic content. Normalized ratios (TCR-N, RCR-N), although improved by Reflection, it still suggests the underutilization of schema and highlights possible avenue for schema-design improvements and better alignment with extracted data.

\subsection{Global Semantic Diversity}

To quantify semantic diversity within the dataset, we analyze the distribution of extracted entities, types, and relationships using information-theoretic measures. Given frequency distributions $\mathbf{p} = \{p_1, p_2, \ldots, p_n\}$ where $p_i$ represents the normalized frequency of element $i$ in the extracted set, we compute Shannon Entropy \cite{shannon1948mathematical}, which provides insights into overall diversity by measuring the balance or skewness of distributions across extracted elements: 

\begin{equation}
H(X) = -\sum_{i=1}^{n} p_i \log_2 p_i
\end{equation} 

In addition, Schema-Normalized Entropy contextualizes diversity within the predefined schema:

\begin{equation}
H_{\text{norm}}(X) = \frac{H(X)}{\log_2 |S|}
\end{equation}
where $|S|$ is the total number of elements defined in the schema. 

Furthermore, Rényi Entropy \cite{renyi1961measures}, with parameter $\alpha = 2$, emphasizes concentration on frequent or rare elements:
\begin{equation}
H_2(X) = -\log_2 \left(\sum_{i=1}^{n} p_i^2\right)
\end{equation}

For each extraction method, we compute these measures on three distributions: 
(1) entity frequencies $\mathbf{p}_E$, (2) entity type frequencies $\mathbf{p}_T$, and (3) relationship frequencies $\mathbf{p}_R$, aggregated across all document chunks.

\begin{table}[htbp]
\renewcommand{\arraystretch}{1.3}
\centering
\caption{Entropy scores across different extraction modes}
\label{tab:diversity_profiling}
\begin{tabularx}{\columnwidth}{l|XXX}
\hline
\textbf{Metric} & \textbf{Single Pass} & \textbf{Multi Pass} & \textbf{Reflection} \\
\hline
\multicolumn{4}{l}{\textit{Shannon Entropy}} \\
\hline
Entity & 7.5383 & 7.2845 & 7.1779 \\
Entity Type & 3.0290 & 2.9835 & 2.8665 \\
Relationship & 5.5438 & 5.6116 & 4.3164 \\
\hline
\multicolumn{4}{l}{\textit{Normalized Schema Entropy}} \\
\hline
Entity Type & 0.6607 & 0.6507 & 0.6252 \\
Relationship & 1.1412 & 1.1552 & 0.8885 \\
\hline
\multicolumn{4}{l}{\textit{Rényi Entropy ($\alpha = 2$)}} \\
\hline
Entity & 2.8619 & 2.5883 &  2.4574 \\
Entity Type & 2.1312 & 2.0851 & 1.9877 \\
Relationship & 3.6355 & 3.5691 & 2.6814\\
\hline
\multicolumn{4}{l}{\textit{Normalized Schema Rényi Entropy ($\alpha = 2$)}} \\
\hline
Entity Type & 0.4648 & 0.4548 & 0.4335 \\
Relationship & 0.7484 & 0.7347 & 0.5520 \\
\hline
\end{tabularx}
\end{table}

\autoref{tab:diversity_profiling} reveals a clear trade-off between extraction completeness and variety. The Reflection method achieves the highest coverage but exhibits the lowest entropy across all dimensions. This result demonstrates that Reflection intentionally reduces diversity to yield a more compact, connected, and navigable graph consistent with predefined extraction rules. Given the significant improvement of Reflection on complementary metrics, this reduction in entropy is well within the acceptable range. Future work will monitor entropy drift and adapt the Reflection rules when diversity falls below the predefined threshold. All extraction methods exhibit moderate schema-normalized entropy, suggesting that the current ontology effectively captures core semantic categories in the corpus, while retaining room for schema enhancement to accommodate greater semantic detail.
\subsection{Comparative Evaluation: LLM-as-a-Judge}
The absence of benchmark ground truth triples for knowledge graph evaluation necessitates an alternative evaluation approach. We leverage LLMs as comparative judges to assess:
\begin{itemize}
\item \textbf{Precision}: Assesses the clarity, specificity, and uniqueness of the extracted triples.
\item \textbf{Faithfulness}: Measures the factual accuracy and grounding of the triples within the source text.
\item \textbf{Comprehensiveness}: Evaluates how completely the generated triples capture the core informational content of the source text. 
\item \textbf{Relevance}: Determines the contextual alignment of triples with the main topics and themes of the source text.
\end{itemize}

This methodology enables ground truth-agnostic metrics that provide relative comparative evaluations rather than absolute measurements, circumventing the need for predefined reference datasets.
According to recent findings \cite{chen2024not, sui2025stop}, Chain-of-Thought (CoT) reasoning \cite{wei2022chainofthought} can underperform in linear, fast, and intuitive tasks, potentially introducing unnecessary complexity and overthinking. Consequently, we adopt a prompting strategy similar to that used by Lopez et al. \cite{lopez2023can}, employing direct instructions without intermediate reasoning steps. In this approach, the model initially commits to a judgment and subsequently provides an explanation. This aligns with cognitive insights indicating that post-hoc rationalization can offer improved human interpretability compared to reasoning that precedes commitment \cite{vamvourellis2025reasoning}.
In our evaluation, we utilized the Qwen3-32B model without reasoning prompts and set the temperature parameter to 0.1.\\

\begin{table}[H]
\centering
\caption{LLM as a Judge Score for various extraction modes}
\begin{tabularx}{\columnwidth}{l|XXX}
\hline
\textbf{Metric (\textuparrow)} & \textbf{Single Pass (\%)} & \textbf{Multi Pass (\%)} & \textbf{Reflection (\%)} \\
\hline
Precision & 22.3 & 38.6 & \textbf{39.1} \\
Faithfulness & \textbf{40.1} & 24.4 & 35.5 \\
Comprehensiveness & 36.3 & 15.6 & \textbf{48.1} \\
Relevance & 34.6 & 28.1 & \textbf{37.3} \\
\hline
\end{tabularx}
\label{tab:llm_judge_results}
\end{table}

To ensure the reliability of our comparative evaluations, we implement a robustness assessment protocol. For each three-way comparison on chunk $c_{ij}$ and metric $\mu$, we conduct $n_{\text{votes}} = 3$ independent evaluations:

\begin{equation}
    \mathcal{V}\mu(c{ij}) = \{J_\mu^{(k)}(c_{ij}, \mathcal{T}{ij}^{(m_1)}, \mathcal{T}{ij}^{(m_2)}, \mathcal{T}{ij}^{(m_3)})\}{k=1}^{3}
\end{equation}

where $k$ represents the $k$-th independent vote. In cases where the votes do not reach consensus, we request a fourth decisive vote.

\begin{table}[h]
\centering
\caption{Consistency across the three LLM as a Judge runs}
\begin{tabularx}{\columnwidth}{l|X}
\hline
\textbf{Metric (\textuparrow)} & \textbf{Agreement (\%)} \\
\hline
Precision         & 82.1 \\
Faithfulness      & 81.3 \\
Comprehensiveness & 86.7 \\
Relevance         & 83.7 \\
\hline
\end{tabularx}
\label{tab:judge_agreement}
\end{table}
\autoref{tab:llm_judge_results} reveals nuanced performance patterns across extraction modes. The reflection mode outperforms other modes in precision, comprehensiveness, and relevance, whereas the single-pass mode excels in faithfulness. This dichotomy underscores a fundamental trade-off between comprehensiveness of extraction and factual grounding.
The reflection mode's relatively lower faithfulness score, despite greater comprehensiveness, suggests increased triple generation potentially exceeds source-constrained accuracy boundaries. 
Moreover, \autoref{tab:judge_agreement} suggests an inverse correlation, where lower agreement is indicative of higher contextual uncertainty.


%% file: sections/discussion.tex
\section{Discussion}

Our comprehensive evaluation framework reveals that the reflection mode delivers the optimal reliability-coverage balance among the three extraction modes. The reflection paradigm demonstrates superior performance across multiple complementary metrics: it achieves the highest CheckRules compliance (64.8\% for all four rules), generates the most triples per chunk (15.8), and wins a clear plurality in LLM-as-a-Judge evaluations across precision, comprehensiveness, and relevance dimensions.

The iterative critic-corrector loop systematically addresses schema violations while expanding triple coverage, yielding a denser yet cleaner knowledge graph. This improvement stems from the reflection mechanism's ability to identify and correct extraction errors while discovering previously missed but valid triples. The reflection mode's superior entity coverage ratio (ECR: 0.53 vs. 0.30-0.31) and relationship coverage ratio (RCR: 0.38 vs. 0.21-0.22) demonstrate its effectiveness in capturing diverse semantic content. The diversity analysis further illuminates the reflection mode's approach: while achieving the highest coverage, it exhibits lower entropy across all dimensions, indicating a deliberate reduction in uncertainty to yield a more compact, connected, and navigable graph.

However, these gains come with some trade-offs. The reflection agent requires additional inference rounds, potentially limiting its suitability for real-time applications requiring fast turnaround (e.g., intraday news feeds). In such scenarios, the single pass strategy may offer a viable alternative, recovering most normalization benefits with reduced computational overhead.

Our evaluation results reveal some limitations. First, cross document co-reference resolution is only partially addressed as the reflection loop operates on isolated filings. Second, our evaluation methodology depends on LLM-voting surrogate ground truth, risking propagation of biases inherent in the underlying judge models. 

%% file: sections/conclusion.tex
\section{Conclusion and Future Work}
In this work, we democratize the access to a reliable financial knowledge graph (KG) triples by releasing an open-source dataset constructed from SEC 10-K filings of all the S\&P 100 companies.
Given the universal and utilitarian nature of SEC 10-K filings, we believe that this open-source dataset will empower developers and researchers to build robust, finance-specific applications, advancing transparency and innovation in the financial domain. 

Our pipeline demonstrates that reliable KG triples can be generated using relatively compact language models (<100B parameters) without costly fine-tuning, provided a robust evaluation and normalization framework is in place—an important consideration for regulated industries where trust and auditability are paramount. This motivated our development of a holistic novel framework for evaluating our KG construction pipeline and it's results. 

We are currently pursuing the below enhancements to further broaden the impact and flexibility of our approach. 

\textbf{Schema-Free KG Construction and Self-Improvement:} Inspired by the Extract-Define-Canonicalize (EDC) paradigm~\cite{zhang2024extract}, we are developing a schema-free pipeline capable of creating schemas from scratch and iteratively refining existing ones. This is particularly valuable for private financial data sources where schema requirements are unknown. Our evaluation results demonstrate that pre-configured schemas may not capture all nuances within financial data, necessitating a schema discovery and improvement pipeline. It would be interesting to compare the results of closed vs. open schema frameworks in terms of coverage and performance degradation. Careful processing \& retrieval of expanded schemas is critical for efficient instruction following during inference.

\textbf{Temporal Knowledge Graphs for Thematic Investing:} Building on recent advances in financial temporal KGs~\cite{li2024findkg}, we plan to enrich our dataset with time-aware features to enable structured event timelines and causal reasoning. Our approach will incorporate temporal annotations capturing the evolution of financial relationships—market events, regulatory changes, and corporate developments—across various time horizons. Temporal KGs can enhance explainability for signal generation for market movements and offer actionable insights for traders and risk managers.

    \textbf{Enhanced LLM-as-a-Judge Evaluation Methodology:} Several factors warrant consideration when interpreting the results of LLM-as-a-judge approach. First, our use of Qwen3-32B without reasoning capabilities may not capture the full potential of more sophisticated reasoning models. Second, the extrinsic validation by using diverse LLM families is necessary to assess cross-model reliability and can help normalize the biases inherent in using a specific model family.
    
\textbf{Table-Aware Serialization for Improved Extraction:}\\
    We observed that smaller language models exhibit conservative extraction behavior from markdown-formatted tables missing semantic relationships despite structured data availability. A dedicated table-aware serialization module could significantly enhance knowledge graph coverage and completeness, particularly for quantitative financial metrics and their inter relationships.

We have used the underlying KG for essential downstream applications, such as question answering, proactive recommendations by integrating real-time news feeds with the structured KG. The promising results highlight the broad applicability of our KG for real-world financial use cases and can be explored at scale. 

Looking ahead, we are making a concerted effort to not only augment our pipeline with above enhancements but also expand our KG triple dataset to include all the S\&P 500 companies with data from the last 10 years of annual SEC 10-K filings, significantly broadening the scope and temporal coverage of our financial KG. We hope to catalyze further research and practical adoption of knowledge graphs in the financial sector and beyond.